\documentclass[conference]{IEEEtran}
\IEEEoverridecommandlockouts
\usepackage{cite}
\usepackage{amsmath,amssymb,amsfonts}
\usepackage{algorithmic}
\usepackage{graphicx}
\usepackage{textcomp}
\usepackage{array}
\usepackage{mdwmath}
\usepackage{mdwtab}
\usepackage{eqparbox}
\usepackage{url}
\usepackage{supertabular}                                
\usepackage{caption}
\usepackage{subcaption}
\usepackage{float}
\usepackage{booktabs}
\usepackage{multirow}
\usepackage{graphicx}
\usepackage{color}
\usepackage{tabularray}
\usepackage{hhline}
\usepackage{wasysym}
\usepackage{rotating}
\usepackage{hyperref}
\usepackage{xcolor}
\usepackage{microtype}
\def\BibTeX{{\rm B\kern-.05em{\sc i\kern-.025em b}\kern-.08em
    T\kern-.1667em\lower.7ex\hbox{E}\kern-.125emX}}
\begin{document}

\title{Towards Weaknesses and Attack Patterns Prediction for IoT Devices
}

\author{
\IEEEauthorblockN{Carlos A. Rivera A., Arash Shaghaghi, Gustavo Batista, Salil S. Kanhere}
\IEEEauthorblockA{School of Computer Science and Engineering, The University of New South Wales (UNSW Sydney), Australia}
}

\maketitle

\begin{abstract}
As the adoption of Internet of Things (IoT) devices continues to rise in enterprise environments, the need for effective and efficient security measures becomes increasingly critical. This paper presents a cost-efficient platform to facilitate the pre-deployment security checks of IoT devices by predicting potential weaknesses and associated attack patterns. The platform employs a Bidirectional Long Short-Term Memory (Bi-LSTM) network to analyse device-related textual data and predict weaknesses. At the same time, a Gradient Boosting Machine (GBM) model predicts likely attack patterns that could exploit these weaknesses. When evaluated on a dataset curated from the National Vulnerability Database (NVD) and publicly accessible IoT data sources, the system demonstrates high accuracy and reliability. The dataset created for this solution is publicly accessible. 

\end{abstract}

\begin{IEEEkeywords}
Weakness (CWE) Prediction, Attack Pattern (CAPEC) Prediction, Pre-deployment Security Check, IoT Security.
\end{IEEEkeywords}

\section{Introduction}
The Internet of Things (IoT) has revolutionised the digital landscape, connecting billions of devices worldwide. While this connectivity brings numerous benefits, it also introduces significant security challenges. Unlike traditional IT devices, IoT devices are often resource-constrained and operate in diverse environments, making them susceptible to a wide range of cyber threats. The integration of IoT into enterprise networks is particularly concerning, as a single compromised device can serve as a gateway for broader network intrusions.

Existing security measures, such as vulnerability scanners and penetration testing, are often inadequate for IoT due to these devices' sheer scale and diversity. Moreover, the constant release of new IoT devices with unique firmware and configurations makes it difficult to maintain a comprehensive security posture. Furthermore, traditional methods for detecting vulnerabilities in IoT devices often necessitate physical access to the device under investigation. These methods, while effective, are limited by the requirement that the device be operational and available for analysis.

To address these challenges, we propose a platform that leverages machine learning to predict potential weaknesses in IoT devices and the attack patterns that could exploit these weaknesses. Our platform utilises a Bidirectional Long Short-Term Memory (Bi-LSTM) model for predicting weaknesses based on textual data related to the device’s characteristics. We then use the Gradient Boosting Machine (GBM) model to predict attack patterns, enabling a proactive approach to IoT security management. Another contribution of this work is the creation of a comprehensive dataset curated from the National Vulnerability Database (NVD) and IoT-specific sources. This dataset includes detailed vulnerability descriptions, associated Common Weakness Enumeration (CWE) categories, device specifications, and attack patterns relevant to IoT environments. The dataset, which is made publicly available, not only underpins the accuracy and reliability of our models but also serves as a valuable resource for future research and development in the field of IoT security.
The remainder of this paper is structured as follows: Section 2 reviews the related work. Section 3 details the proposed solution and reports on the implementation and evaluation results. Section 4 concludes the paper and suggests directions for future research. The link to the publicly available datasets created for this project is available in Section 5.

\section{Related Work}

Several approaches have been proposed in the literature to address the vulnerabilities and security risks associated with IoT devices. This section reviews the existing work on IoT cybersecurity, with a particular focus on machine learning-based methods, natural language processing (NLP) techniques, and the integration of these methods for weaknesses and attack pattern prediction.

Artificial intelligence has increasingly been applied to various aspects of cybersecurity, including malware detection, vulnerability analysis, and penetration testing. Kotenko et al. \cite{DBLP:journals/access/KotenkoSB18} proposed a malware detection system that leverages machine learning algorithms to analyse traffic data from IoT devices. Their approach utilised five different machine learning algorithms to process large volumes of traffic data, ultimately relying on a voting mechanism to enhance accuracy. However, the framework was limited to binary classification and heavily depended on filtering out non-IoT device traffic, which poses challenges in mixed network environments.

Fang et al. \cite{Fang2019TAPAS} introduced a static code analysis model for penetration testing, using an LSTM-based approach to analyse PHP source code. While their model successfully identified several weaknesses, it struggled with distinguishing safe code from unsafe code in certain instances. This limitation highlights the challenges of using AI for precise weakness detection in IoT environments, where the diversity of code and configurations can be vast.

Li et al. \cite{DBLP:journals/access/LiFZT19} advanced vulnerability analysis through the use of a lightweight deep neural network, specifically a Bidirectional LSTM (Bi-LSTM), to classify functions as vulnerable or non-vulnerable. Although their model showed promise in binary classification tasks, it did not address the more complex multi-class classification problem, which is often necessary in the context of IoT vulnerabilities where multiple types of vulnerabilities may coexist.

NLP techniques have gained traction in cybersecurity for processing and analysing textual data related to vulnerabilities and threats. Saletta and Ferretti \cite{DBLP:conf/dasc/SalettaF20} utilised NLP to map source code into an Abstract Syntax Tree (AST) and employed machine learning classifiers to identify software weaknesses. Their use of the Doc2Vec embedding method allowed for effective classification of Java code vulnerabilities, though the reduction in the number of classes from 25 to 8 during data analysis suggested a trade-off between model complexity and performance.

Wåreus et al. \cite{DBLP:conf/icissp/WareusDTH22} proposed a Hierarchical Attention Network with Virtual Adversarial Training to classify security-related issues from GitHub. Their approach combined term frequency with word embeddings to enhance the model’s understanding of the context, particularly in documents with varying levels of technical detail. However, the model’s performance was limited when applied to datasets with significant vocabulary differences, such as those between technical and community-generated content.

Ziems and Wu \cite{DBLP:journals/corr/abs-2105-02388} explored the use of BERT Transformers and LSTM models for vulnerability classification. Their comparison highlighted the strengths of Transformer models in handling complex NLP tasks, although the absence of certain evaluation metrics, such as confusion matrices, made it difficult to fully assess the model’s predictive capabilities. Their work underscored the importance of selecting appropriate NLP techniques to manage the unique challenges posed by cybersecurity text data.

In the context of IoT devices, several studies have explored the use of machine learning to predict weaknesses and potential attack patterns. Anthi et al. \cite{DBLP:journals/iotj/AnthiWSTB19} developed a system that processes traffic from IoT devices using multiple machine learning algorithms to classify devices, detect attacks, and identify the type of attack. While their approach demonstrated high accuracy, it was limited to a small number of device types and attack scenarios, which may not scale effectively in more diverse IoT environments.

Zhou et al. \cite{DBLP:conf/infocom/ZhouHLHW18} extended this work by pre-training a deep learning model on a large dataset before applying it to IoT traffic data. Their approach showed strong results in binary classification of attack types but was limited in its ability to generalise to a broader range of attack vectors, reflecting a common challenge in IoT security research.

Blinowski and Piotrowski \cite{10.1007/978-3-030-48256-5_9} focused on classifying vulnerabilities specific to IoT hardware using a Support Vector Machine (SVM). Their model achieved moderate success but struggled with accurately predicting less common vulnerability types, a challenge often encountered when dealing with imbalanced datasets in cybersecurity.

Nui et al. \cite{niu2020deep} proposed a deep learning-based static taint analysis method to identify software vulnerabilities in IoT devices, focusing on two specific weaknesses. While their approach achieved high accuracy, it was limited to a small subset of vulnerabilities, highlighting the need for more comprehensive models that can address the wide variety of threats faced by IoT devices.

Rivera et al. \cite{DBLP:conf/mobiquitous/AlvarezSNK21} introduced a multi-class classification framework using Gradient Boosting Decision Trees (GBDT) to predict the risk label associated with the risk score of vulnerabilities in IoT devices. Their model showed promise with an accuracy of 71\%, though it did not fully leverage the potential of NLP techniques in conjunction with machine learning for a more robust prediction framework.

Recent advancements have seen the integration of NLP and transfer learning in IoT security to enhance predictive capabilities. Wang et al. \cite{DBLP:conf/cse/WangQC21} and Shankar et al. \cite{DBLP:conf/dsaa/DasSHPA21} both employed BERT-based models to classify vulnerabilities using weakness descriptions (CWEs) as labels. These models utilised advanced embedding techniques to improve classification accuracy, though they were often challenged by the need for large, well-balanced datasets.

Fatih and Hwang \cite{DBLP:conf/IEEEcloud/BulutH21} developed the NL2Vul framework, which uses transfer learning and NLP to predict vulnerability scores and severity from textual descriptions. Their multi-class classification approach achieved competitive results but was hindered by the inherent imbalance in vulnerability datasets, a common issue in IoT security research.

The work of Kuppa et al. \cite{DBLP:conf/IEEEares/KuppaAL21} represents a notable attempt to generate attack labels from vulnerability descriptions using a BERT Transformer, although their reliance on specific phrases to determine attack types introduced potential inaccuracies. This work emphasises the importance of accurate label generation and the challenges posed by the unsupervised nature of some NLP tasks.

\section{Proposed Solution}

The proposed platform is designed to enhance the security of IoT devices by predicting potential weaknesses and associated attack patterns. Given the heterogeneity and complexity of IoT environments, traditional security measures often fail to provide comprehensive protection. Our platform addresses this gap by integrating two machine learning models that operate in tandem: a Bidirectional Long Short-Term Memory (Bi-LSTM) network for predicting weaknesses and a Gradient Boosting Machine (GBM) for predicting attack patterns. This dual-model approach ensures a holistic security assessment by identifying both weaknesses and the related attack patterns that could be used to exploit them.

\subsection*{\textbf{Weakness Prediction Framework}}

The methodology's core lies in the Weakness Prediction Framework, which is designed to predict specific weaknesses in IoT devices based on textual descriptions of vulnerabilities. This framework processes information associated with IoT devices through syntactic and semantic analysis, thereby determining potential weaknesses that malicious entities may exploit. The Weakness Prediction Framework is centred around a Bidirectional Long Short-Term Memory (Bi-LSTM) network, a recurrent neural network (RNN) that is particularly effective for processing sequential data, such as text. This framework analyses the textual information related to IoT devices and predicts potential weaknesses based on historical vulnerability data.

\subsection{Dataset for Weakness Prediction Framework}

In constructing the datasets for this study, we paid particular attention to common pitfalls in data acquisition, such as limited diversity, missing labels, and imbalanced data distributions. To mitigate these issues, we sourced data from three distinct repositories: the IoT dataset used by Rivera et al., the National Vulnerability Database (NVD), and technical data from the ZoomEye search engine.

The datasets were carefully curated to include various features, such as the brand of the device, product type, model, operating system, and detailed descriptions of vulnerabilities. These features were compiled to represent IoT devices and their associated weaknesses. The dataset was further enriched by merging structured data from Rivera et al. \cite{DBLP:conf/mobiquitous/AlvarezSNK21} with unstructured text data from the NVD and ZoomEye.

The resulting datasets were categorised as “Only-IoT” and “All-Systems.” The Only-IoT dataset focuses exclusively on IoT devices, while the All-Systems dataset includes a broader range of systems, providing a more diverse training set for the models. Both datasets were subjected to rigorous preprocessing steps, including data cleaning and normalisation, to ensure the information was accurate and ready for analysis.

We encountered two primary challenges during dataset construction. First, the information in the initial dataset was structured and related primarily to vulnerabilities rather than weaknesses. We addressed this by combining textual features and matching them with corresponding vulnerability codes from the NVD, thus allowing us to extract the relevant weaknesses. Second, we supplemented the technical details of each device by querying ZoomEye and appending the information to the dataset, enhancing the robustness of the data.

The datasets were then expanded by adding records from the NVD, ensuring that the dataset was sufficiently large to support the training of deep learning models. After proper sanitisation and elimination of records with undefined weaknesses, the final dataset sizes were 4,892 for the Only-IoT dataset and 75,559 for the All-Systems dataset.

To address the issue of class imbalance, we undertook several steps. First, we analysed the distribution of records across classes and eliminated classes with fewer than 30 records. This reduced the impact of outliers and helped balance the dataset. Second, we generated additional copies of the dataset, further balancing the distribution of records among classes. Lastly, we utilised a hierarchical structure to group related weaknesses, reducing the number of distinct classes and helping the model focus on predicting a smaller set of grouped labels. The dataset, including comprehensive details on its content, can be accessed through the project’s GitHub repository at the following link: \url{https://github.com/criveraalvarez/IoT_CWE-CAPEC_Dataset}.

\subsection{Model Selection}

We selected a combination of Recurrent Neural Networks (RNN) and pre-trained transformer models to implement the weakness prediction framework. Specifically, we employed the Bidirectional Long Short-Term Memory (Bi-LSTM) model and three transformer models: BERT, BART, and Longformer. These models were chosen for their proven effectiveness in NLP tasks and their ability to capture complex syntactic and semantic relationships within textual data.

The Bi-LSTM model was employed as a baseline due to its ability to process sequences in both forward and backward directions, capturing contextual information more effectively than traditional RNNs. This model addresses the “vanishing and exploding” gradient problems typically associated with conventional RNNs. Additionally, we implemented word embeddings, such as Word2Vec, to enhance the model’s understanding of the relationships between words in the text.

The transformer models, on the other hand, leverage attention mechanisms to generate more accurate predictions by focusing on the most relevant parts of the input text. These models have revolutionised the field of NLP by introducing the attention module, which allows for processing input sequences in parallel rather than sequentially. This results in significantly faster training times and improved performance on various NLP tasks.

Given the high computational demands of training transformer models from scratch, we used pre-trained versions of BERT, BART, and Longformer. These models were fine-tuned on our datasets using transfer learning, which involves adjusting the weights of a pre-trained model to suit a specific task. Transfer learning is particularly advantageous in NLP, where large-scale pre-training on massive corpora can be leveraged to achieve superior performance on smaller, domain-specific datasets.

The preprocessing steps involved cleaning the text data, tokenising it into numerical representations, and applying embeddings to capture the semantic relationships between words. For the Bi-LSTM model, we used pre-trained Word2Vec embeddings, while the transformer models utilised their built-in embedding mechanisms. This allowed us to take full advantage of the contextual information embedded in the text, leading to more accurate predictions.

\subsection{Experimental Setup \& Implementation}

The experimental setup was designed to ensure robust model training and evaluation. We utilised the Google Colab environment and the high-performance computing resources provided by the University of New South Wales (UNSW). This service comes from a server with four Tesla T4 GPU processors and fifteen Gigabytes (15GB) in memory per processing unit. The datasets were split into training and testing sets, with 90\% of the data allocated for training and 10\% for testing. Additionally, we applied k-fold cross-validation to improve model performance and reduce overfitting.

For the Bi-LSTM model, training was conducted using the PyTorch framework, with early stopping implemented to prevent overfitting. The model was trained for varying epochs (100, 500, 800, and 1000) to determine the optimal training duration. For the transformer models, fine-tuning was carried out using the HuggingFace library, with each model fine-tuned over 100 epochs.

To accommodate the higher computational requirements of the Longformer model, we used the Colab Pro+ service, which provides access to additional GPU resources and increased memory capacity. This allowed us to efficiently fine-tune the Longformer model, which is particularly well-suited for processing long documents due to its ability to handle extended input sequences.

The training and testing processes were conducted using cloud-based environments, ensuring that the models could be trained and evaluated in a scalable and reproducible manner. The data was prepared using a set of Python libraries, including NumPy, Pandas, scikit-learn, Seaborn, and Matplotlib, which facilitated data manipulation, analysis, and visualisation.

\subsection{Evaluation Results}

We used evaluation metrics for classification such as Precision, Recall, F1-Score and Accuracy \cite{DBLP:journals/corr/abs-2008-05756}. For instance, Precision measures the proportion of predicted values as true positives. Recall measures the fraction of true positives, including true positives and false negatives. These metrics feed the harmonic mean (F1-Score). This metric assesses the model's capability to analyse Precision and Recall, then assigns larger weights to smaller classes but rewards models that provide similar Precision and Recall values.

The metrics for classification vary depending on the type of classification, binary or multi-class. The current implementation fits the multi-class type since the predicted label contains different weaknesses. This approach calculates the metrics using the micro and macro approaches. The micro approach assigns weights to each class, and the macro assigns equal weights to all the classes. We use the micro approach as its results provide more granular evaluation results.  
Further, in a multi-class approach, the results from F1-score and accuracy produce equal values(Equations in \ref{metrics-multiclass}). 

\begin{equation} \label{metrics-multiclass}
\scalebox{0.85}{$ 
\begin{split}
Precision_{k}=\frac{TP_{k}}{TP_{k}+FP_{k}}
\quad Recall_{k}=\frac{TP_{k}}{TP_{k}+FN_{k}} \\\\
Precision_{Micro}=\frac{\sum_{k=1}^{K}TP_{k}}{\sum_{k=1}^{K}Total Column_{k}}
=\frac{\sum_{k=1}^{K}TP_{k}}{Grand Total} \\\\
Recall_{Micro}=\frac{\sum_{k=1}^{K}TP_{k}}{\sum_{k=1}^{K}Total Row_{k}}
=\frac{\sum_{k=1}^{K}TP_{k}}{Grand Total} \\\\
F1-score_{Micro}=\frac{\sum_{k=1}^{K}TP_{k}}{Grand Total}=Accuracy \\\\
\end{split}
$}
\end{equation}

Let k represent any class from K, TP the true positives, FP the false positives, and FN the false negatives.\\

We implemented the RNN Bi-LSTM models and pre-trained transformer models Bert, Bart, and Longformer. All carry out the text classification task, fitting the multi-class approach. To compare the performance of all models, we use the Friedman test from   \cite{DBLP:journals/jmlr/Demsar06}. Here, the author recommends non-parametric tests to compare classifiers statistically. We follow the Friedman test, which ranks the performance results from different models over multiple datasets (two datasets as a minimum). This test will assign rank-1 to the best model and rank-2 to the second best, and so on. In ties, average ranks are assigned. Later, the test calculates the average for each model and all the datasets. Lastly, the Friedman statistic (Equation \ref{metrics-friedman}) and Iman-Davenport Statistic (Equation \ref{metrics-ImanDavenport}) are calculated. The former is sufficient for problems with ${k>5}$ and ${N>10}$; however, our implementation is smaller, and thus we use the latter. The Iman-Davenport test produces a number that can be compared with critical values for the Friedman test\cite{martin1993tables}. The null hypothesis (H$_{0}$) can be rejected if the result is more significant than the corresponding critical value. In our case, the H$_{0}$ states that all the models are equivalent, and so are their ranks.\\

\begin{equation} \label{metrics-friedman}
\scalebox{0.85}{$
\begin{split}
Friedman = \chi^{2}_{F}= \frac{12N}{k(k+1)} \left[ \sum_{j} R^{2}_{j} - \frac{k(k+1)^{2}}{4} \right]
\end{split}
$}
\end{equation}

\begin{equation} \label{metrics-ImanDavenport}
\scalebox{0.85}{$
\begin{split}
Iman\text{-}Davenport = F_{F} = \frac{(N-1)\chi^{2}_{F}}{N(k-1)-\chi^{2}_{F}}
\end{split}
$}
\end{equation}

We exhibit the accuracy score(test dataset) results from running ten epochs of the Bi-LSTM model and each pre-trained Transformer model (BERT, Longformer and BART) in Table \ref{table:ACC_Comparison_V}. 
To analyse the models' results, we generated the ranking (using Friedman Ranking) for each model result, and at the bottom, we calculated the mean rank for each model. The results are disclosed in Table \ref{table:Friedman}. We used these results with the Iman-Davenport test(F$_{F}$) to determine whether to accept or reject the null hypothesis (That the models are equivalent, so are their ranks). The test results are shown in Table \ref{table:Iman-DAvenport_results}. We note the following: 1) The Iman-Davenport test confirms the null theory (all the models are equivalent) and will produce similar results, 2) The Bi-LSTM model achieved the highest accuracy in two of the six datasets, 3) The Longformer Transformer model achieved the utmost accuracy in three of the six datasets, 4) The Bart Transformer model achieved the greatest accuracy only in one of the six datasets, 5) The Bert Transformer model was the only Transformer model without top accuracy results, 6) The mean ranking from all the models confirmed the Longformer Transformer model as the best amongst all the models tested. (The smaller value in Friedman's mean rank shows the best-performing model.)

\begin{table}
\caption{Comparison of the models' Test Accuracy$_{micro}$. The models are trained over ten epochs; the labels used to identify the datasets are "AS" to represent the All-Systems dataset and "OI" to represent the Only-IoT dataset. Bold values are the top among the models and per dataset.}
\label{table:ACC_Comparison_V}
\centering
\begin{tabular}
{|m{1.3cm} | m{1.4cm} | m{1.2cm} | m{1.1cm} | m{1.5cm} |}
\hline
\textbf{Dataset} &\textbf{Bi-LSTM} &\textbf{BERT} &\textbf{BART} &\textbf{Longformer}\\
\hline 
OI V1.1     &48.1\%         &46.7\%     &49.7\%          &\textbf{50.4\%}\\
\hline 
OI V1.2     &\textbf{49.6\%}&39.7\%     &43.0\%          &40.6\%\\
\hline 
OI V1.3     &28.4\%         &24.5\%     &27.3\%          &\textbf{29.9\%}\\
\hline 
AS V2.1     &\textbf{79.1\%}&76.1\%     &76.3\%          &76.4\%\\
\hline 
AS V2.2     &67.9\%         &70.3\%     &36.4\%          &\textbf{70.6\%}\\
\hline 
AS V2.3     &40.5\%         &49.8\%     &\textbf{49.9\%} &48.7\%\\
\hline 
\end{tabular}
\end{table}

\begin{table}
\caption{Comparison of the models' Friedman Ranking. The datasets' labels are "AS", which represents All-Systems DS, and "OI", which represents Only-IoT DS.}
\label{table:Friedman}
\centering
\begin{tabular}
{|m{1.5cm} | m{1.3cm} | m{1cm} | m{1cm} | m{1.6cm} |}
\hline
\centering \textbf{Dataset} &\textbf{Bi-LSTM} &\textbf{BERT} &\textbf{BART} &\textbf{Longformer}\\
\hline 
OI V1.1             &4     &3       &2         &1\\
\hline 
OI V1.2             &1     &4       &2         &3\\
\hline 
OI V1.3             &2     &4       &3         &1\\
\hline 
AS V2.1             &1     &4       &3         &2\\
\hline 
AS V2.2             &3     &2       &4         &1\\
\hline 
AS V2.3             &4     &2       &1         &3\\
\hline
Mean Rank           &2.46  &2.83    &2.65      &2.36\\
\hline
\end{tabular}
\end{table}

Once we confirmed that all the models would produce similar results, we conducted a more profound analysis of their performance, looking to improve their performance results. We verified all the models' accuracy, training loss and validation loss. The goal is to check whether the models were overfitting. The models' results showed the same behaviour: 1) No model was overfitting, 2) The accuracy is growing but not yet stabilised, and 3) Both loss functions were improving but not yet stabilised.

\begin{table}
\caption{Result of computing the Iman-Davenport Test (F$_{F}$) over all the models and datasets and the final resolution about the null hypothesis (H0).}
\label{table:Iman-DAvenport_results}
\centering
\begin{tabular}
{|m{1.5cm} | m{2cm} | m{3cm} | }
\hline
\textbf{F$_{F}$} &\textbf{Critical Value} &\textbf{Accept/Reject H0}  \\
\hline 
0.07129          &7.6      & F$_{F}$ < Critical Value $\therefore$ Do not reject H0.\\
\hline
\end{tabular}
\end{table}

The results from the 100 epochs re-run are provided in Table \ref{table:ACC_Comparison_V100}, and our findings are as follows: 
\begin{itemize}
\item{The Longformer transformer model is still the top-ranked.}
\item{All the models' accuracy improved by at least 15\%, moving the maximum accuracy to 65.4\%.}
\item{The losses in all the models show that no model showed overfitting behaviour.}
\item{The accuracy metric behaviour in any model showed that they still need to run for more epochs to stabilise.}
\end{itemize}

\begin{figure*}[t]
    \includegraphics[width=0.8\textwidth]{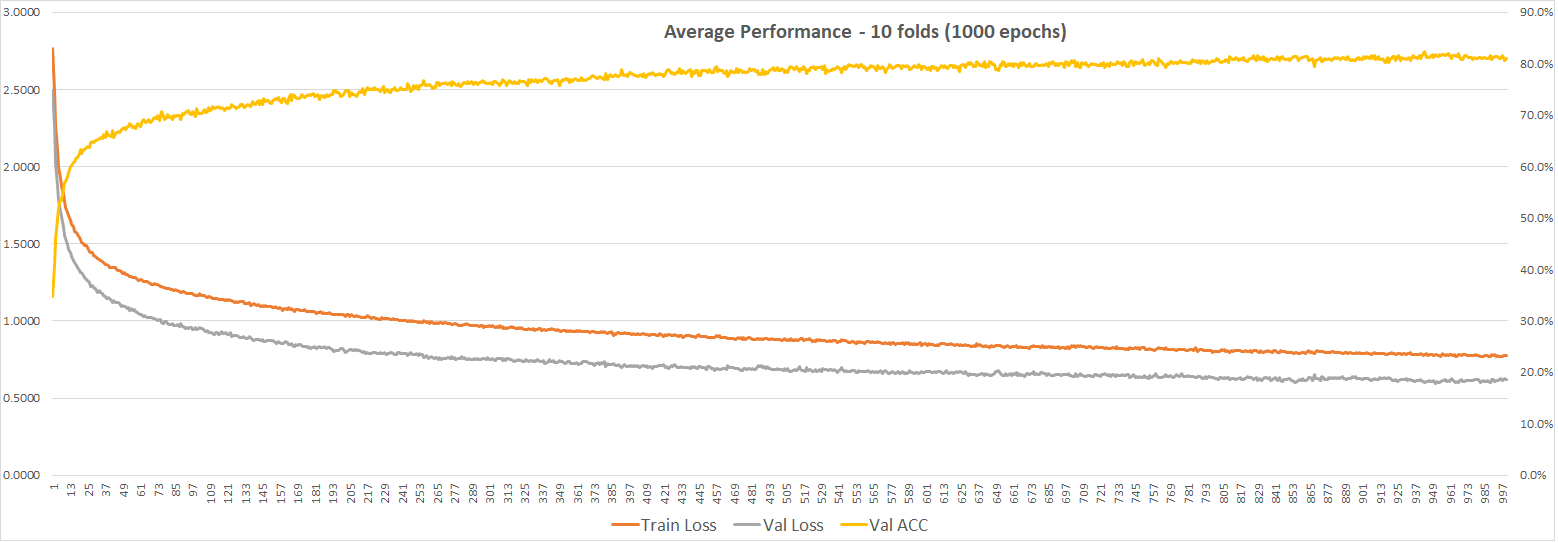}
    \centering
    \caption{Accuracy, Train Loss, and Test Loss from training 1K-epochs the Bi-LSTM model over the OI V1.1 dataset.}
    \label{fig:ACC-1K-epochs}
\end{figure*}

The result from this last re-training (Figure \ref{fig:ACC-1K-epochs}) reveals that all the models still need to be trained for more epochs to start stabilising (The accuracy and loss functions need to indicate the same results for several epochs). For this extended training, we selected the model that consumes less memory and computation resources, the Bi-LSTM. We re-ran the training for 500, 800, and 1000 epochs. Test results are shown in Table \ref{table:1K_epochs}.   

We plotted (Figure \ref{fig:ACC-1K-epochs}) the 1K-epochs performance results Accuracy$_{Micro}$, train and test loss function values from the Bi-LSTM model which used the OI V1.3 dataset.
We found that the Bi-LSTM model showed stabilisation after the 800 epoch. Nevertheless, after completing the 1k-epochs training, the model still has no signs of overfitting. 
We conclude that the Bi-LTSM model can predict weaknesses in IoT devices with an accuracy of 77\% (though capable of running for a couple of hundred more epochs and achieving more than 80\% accuracy).

\begin{table}
\caption{Comparison of the models' Test Accuracy$_{micro}$. The models are trained over 100 epochs using only the Only-IoT datasets V1.1, and all the models have enabled the early stopping function. Bold values are the top among the models and per dataset.}
\label{table:ACC_Comparison_V100}
\centering
\begin{tabular}
{|m{1.5cm} | m{1.4cm} | m{1.1cm} | m{1.1cm} | m{1.6cm} |}
\hline
\textbf{Metric} &\textbf{Bi-LSTM} &\textbf{BERT} &\textbf{BART} &\textbf{Longformer}\\
\hline 
Accuracy                &64.2\%         &63.9\%         &64.6\%     &\textbf{65.4\%}\\
\hline 
Train Loss              &1.3732         &0.4919         &0.4467     &0.4467         \\
\hline 
Test Loss               &1.2346         &0.4601         &0.3965     &0.3965         \\
\hline 
Rank                    &3              &4              &2          &1              \\
\hline
\% change from 10e      &16\%           &17\%           &15\%       &15\%           \\
\hline
\end{tabular}
\end{table}

\begin{table}
\caption{Accuracy results of computing the Bi-LSTM for 500, 800 and 1000 epochs.}
\label{table:1K_epochs}
\centering
\begin{tabular}
{|m{1.4cm} |m{1.4cm} | m{1.4cm} | m{1.3cm} | m{1.2cm} | }
\hline
\textbf{Metric} &\textbf{100 epochs} &\textbf{500 epochs}& \textbf{800 epochs} &\textbf{1K epochs}  \\
\hline 
Accuracy        &64.2\%      &74.0\%      &76.0\%  &77.0\%  \\
\hline
Train Loss      &1.3732      &1.0657      &0.9813  &0.9430  \\
\hline
Test Loss       &1.2346      &0.8518      &0.7792  &0.7474  \\
\hline
\end{tabular}
\end{table}


\begin{figure}
\includegraphics[width=0.5\textwidth]{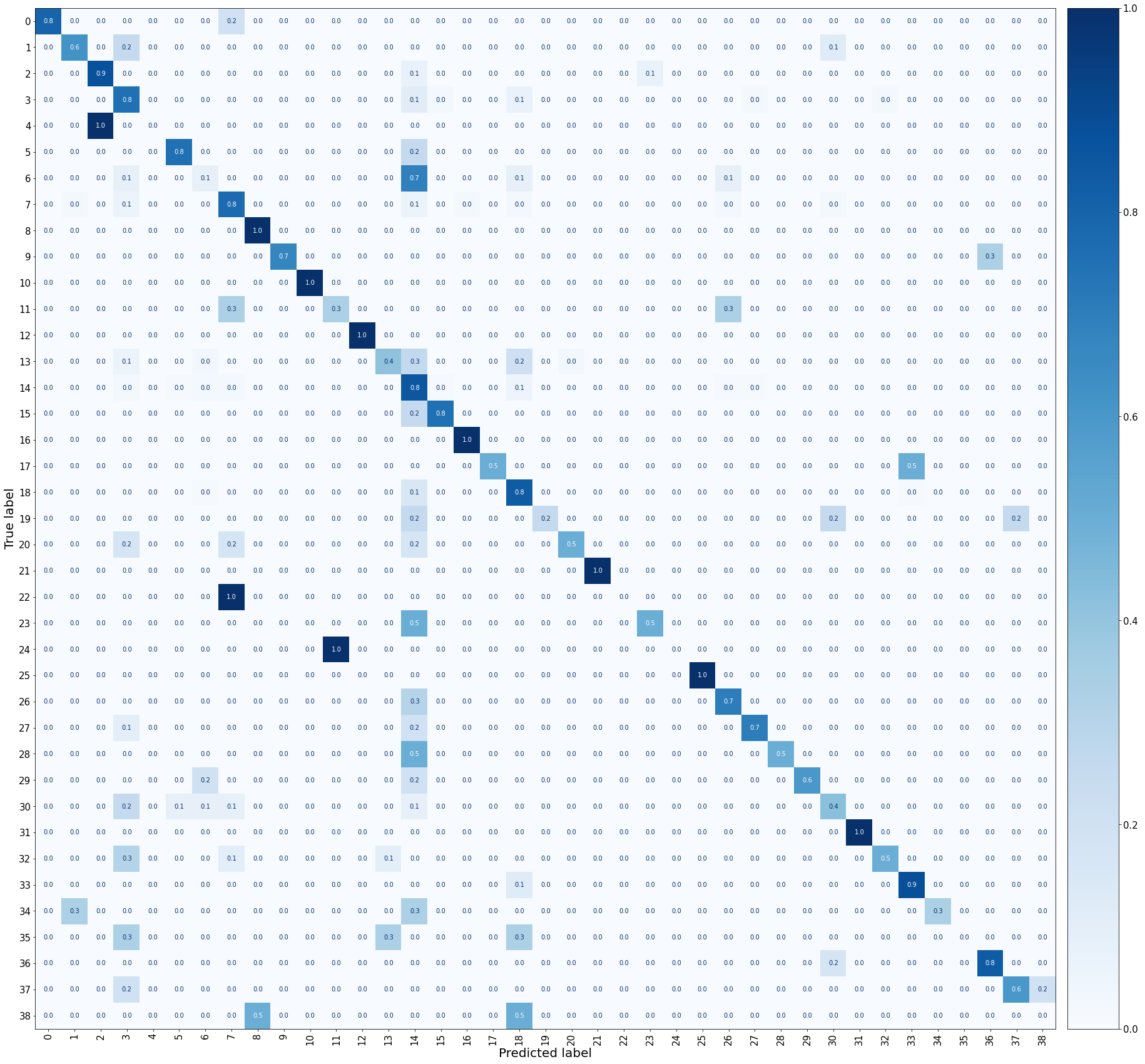}
\centering
\caption{Confusion matrix from the Bi-LSTM model and from loading the 1K epoch on the OI V1.1 dataset test.}
\label{fig:CM15}
\end{figure}

We use the confusion matrix to ascertain how the classes are predicted or confused. From this method, we can determine the probability of each class being correctly or incorrectly predicted. The confusion matrix plot from the Bi-LSTM model and processing the test datasets V1.1 is available in Figure \ref{fig:CM15}).

The confusion matrix and the comparison tests enabled us to conclude the following:
\begin{itemize}
\item{The Bi-LSTM model greatly improved from reaching an accuracy metric of 77\%.}
\item{The Bi-LSTM confusion matrix shows the model struggles with six classes; four of them completely confuse the correct class with another, and in the other two cases, the model confuses with high probabilities the incorrect classes.}
\item{In general, the confusion matrix (Figure \ref{fig:CM15}), showed the Bi-LSTM model at its 1K-epoch has identified most of the classes with high accuracy.}
\item{The Bi-LSTM model performance after training over a thousand epochs demonstrated no need for new technologies such as the pre-trained transformers.}
\item{We confirm the importance and the positive effect in the results of any Bi-LSTM model and Pre-trained transformer models when training with a large enough dataset such as the OI V1.1.}
\item{The models will perform poorly if the dataset size is shorter, despite the better balance of the classes (we tested with two reduced in size and more balanced datasets for each primary dataset).}
\end{itemize}

\subsection{Multi-Class Multi-Label Implementation}\label{multilabel}
After completing the multi-class model, we implemented the multi-class, multi-label model. Here, we provide our conclusions: 
First, we trained and tested the model over the AS Dataset. However, this version of the DS contains more classes for the model to predict. While the multi-class version predicts 46 classes, this model predicts 57 classes. 
Second, the multi-class/label model achieves metrics similar to the multi-class version's. Specifically, the F1-Score is almost equal; however, the accuracy obtained is slightly lower, an effect directly linked to the increase of classes.
Third, the vocabulary in this implementation was not increased, as we seek to compare it with its simile, which does not have added words. The model metrics obtained are better with the addition of specialised vocabulary. 
Fourth, this model version provides a more complex result than the multi-class version, capable of predicting more than one label. However, the lack of enough samples related to the second label forced our decision to keep the multi-class solution as the more adequate. 

In table \ref{table:MultiClassLabel}, we compare the metric results from implementing multi-class/one-label and multi-class/multi-label BERT, BART and Longformer transformer models, trained and tested over the AS DS.

\begin{table}[h!]
\caption{Results from the multi-class/one-label and multi-class/multi-label BERT, BART and Longformer transformer models, trained and tested over the AS dataset.}
\label{table:MultiClassLabel}
\centering
\begin{tabular}
{|m{1.6cm} |m{1.2cm} | m{1.1cm} | m{1.1cm} | m{1.2cm} | }
\hline
\textbf{Model} &\textbf{Precision} &\textbf{Recall} &\textbf{F1-Score}  &\textbf{Accuracy}\\
\hline 
BERT One-Label         &77\%             &77\%             &77\%               &77\%\\
\hline 
BERT Multi-Label       &82\%             &72\%             &76\%               &75\%\\
\hline 
BART One-Label         &76\%             &77\%             &77\%               &76\%\\
\hline
BART Multi-Label       &84\%             &72\%             &77\%               &76\%\\
\hline
Longformer One-Label   &76\%             &77\%             &77\%               &77\%\\
\hline
Longformer Multi-Label &82\%             &73\%             &77\%               &76\%\\
\hline
\end{tabular}
\end{table}

\subsection*{\textbf{Attack Patterns Prediction Framework}}
\setcounter{subsection}{0}

The conception of the Attack Patterns Prediction Framework began with an analysis of the relationships among IoT devices, their associated weaknesses, and potential attack patterns. Our primary references were the cybersecurity databases maintained by MITRE and the National Institute of Standards and Technology (NIST), specifically the National Vulnerabilities Database (NVD). The NVD provides records linked to one or more weaknesses, each manually associated with one or more attack patterns. These attack patterns are cataloged in the Common Attack Pattern Enumeration and Classification (CAPEC) database, which MITRE organises into a hierarchical tree structure. The CAPECs are classified into nine main classes within the “Mechanisms of Attack” view, the most commonly used by researchers to understand attack strategies.

Through this hierarchical analysis, we gained a comprehensive understanding of the CAPECs. However, it became apparent that the CAPECs classification was more suited for understanding attacks in a general sense than for direct application in a predictive model. Consequently, we adopted a more systematic approach to classifying the CAPECs by utilising the Advanced Persistent Threat (APT) concepts and the Kill Chain. The APT framework involves multiple attackers targeting specific enterprise-level organisations, with the attack process divided into stages, each representing both an opportunity for the attacker to advance and for the defender to intercept.

\begin{figure*}[t]
\includegraphics[width=0.7\textwidth]{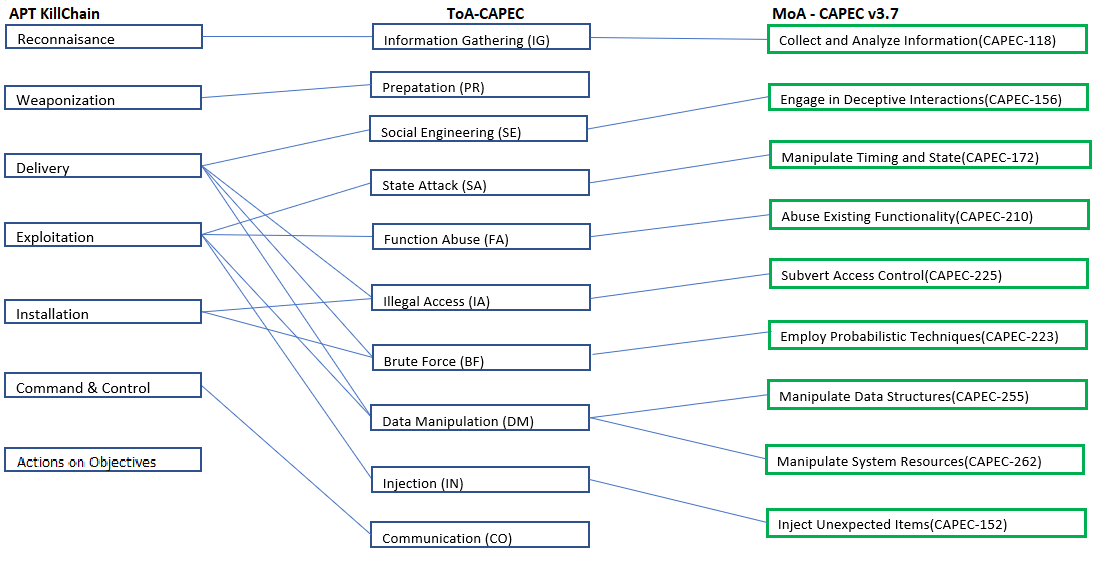}
\centering
\caption{Mapping between APT Kill Chain, Types of Attack Classes, and CAPECs from Mechanisms of Attack View.}
\label{fig:APT-CAPEC}
\end{figure*}

Building on the classification method used in prior research, we categorised the CAPECs into the following main classes: Information Gathering, Injection, Social Engineering, State Attack, Function Abuse, Brute Force, Illegal Access, and Data Manipulation. Additionally, two classes, Preparation and Communication, were identified as critical stages in the APT Kill Chain, although no current CAPECs directly map to these stages. Preparation involves the initial steps an attacker takes before launching an attack, while Communication refers to establishing control channels with external servers, commonly known as Command and Control (C\&C). Figure \ref{fig:APT-CAPEC} illustrates the mapping between the APT Kill Chain, the attack types, and the corresponding CAPECs.

Our framework uses this CAPEC mapping as a label to predict attack patterns for IoT devices with identified weaknesses. This structured approach enables us to systematically predict how attackers might exploit specific weaknesses in IoT devices, thereby providing actionable insights for strengthening security before deployment.

\subsection{Dataset for Attack Patterns Prediction Framework}

We constructed the dataset for our prediction framework by aggregating data from various sources, building upon the foundation established in \cite{DBLP:conf/mobiquitous/AlvarezSNK21}. The elements and corresponding sources are detailed in Table \ref{table:CAPEC_DS}. To enhance the dataset, we incorporated weaknesses associated with each IoT device. It is important to note that each record in this dataset is linked to multiple vulnerabilities, as catalogued by the National Vulnerability Database (NVD). Due to these associations, the dataset contains repeated records; however, each instance varies based on the type and number of weaknesses (Common Weakness Enumerations, CWEs) associated with each vulnerability. For the purpose of the current prediction task, it was necessary to establish a link between the devices, their weaknesses, and the Common Attack Pattern Enumeration and Classification (CAPEC) categories. This challenge was addressed using the mapping methodology proposed in \cite{DBLP:journals/virology/LuhTTSJ20}. The final structure, along with detailed descriptions of each feature in the constructed dataset, is presented in Table \ref{table:CAPEC_DS}.

\begin{table*}[h!]
\caption{Description of features we gathered to create the CAPECs dataset we use to predict the CAPECs in IoT devices. We show the details for each feature. The sources labels are: SRC-1 from  \cite{DBLP:conf/mobiquitous/AlvarezSNK21}, RC-2 from NVD, and SRC-3 from \cite{DBLP:journals/virology/LuhTTSJ20}.}
\label{table:CAPEC_DS}
\centering
\begin{tabular}
{|p{1.3cm} |p{3cm}     |p{2.3cm}               |p{1.9cm}           |p{5.3cm}|}
\hline
\textbf{Source}& \textbf{Feature Name}  &\textbf{Data Type} &\textbf{Unique Values} &\textbf{Details}\\
\hline 
SRC-1 &Brand        &Categorical    &129     &Name of the device reported on the CVE.\\
\hline 
SRC-1 &Product Type &Categorical    & 71     &Phrase describing the product.\\
\hline
SRC-1 &Category     &Categorical    &5      &SmartHome, Medical, Wearable, Telecomm, and Other.\\
\hline
SRC-1 &Price        &Continuous     &Infinite &Reported in US Dollars.\\
\hline
SRC-1 &Year Difference &Continuous &Infinite &Difference of Years between a vulnerability was reported and the product release year.\\
\hline
SRC-1 &Protocols    &Categorical    &8      &Protocols used in Communication Capability.\\
\hline
SRC-1 &Data Storage &Binary         &2      &Location of data Locally or Remotely.\\
\hline
SRC-1 &Personal Information &Binary &2      &Tracks physical location: Yes or No.\\
\hline
SRC-1 &Communication Capability &Categorical &31 &Communication technology.\\
\hline
SRC-1 &Authorisation Encryption &Categorical &4 &Encryption used: Symmetric, Asymmetric, None, or Both.\\
\hline
SRC-2 &Weakness ID-1 &Categorical &9 &From NVD, identification code of each weakness(CWE) associated per vulnerability.\\
\hline
SRC-2 &Weakness ID-2 &Categorical &9 &From NVD, identification code of each weakness(CWE) associated per vulnerability.\\
\hline
SRC-3 &Associated CAPEC-IDs [8 Class Labels] &Binary &2 &Mapped APT-CAPEC categories list.\\
\hline
\end{tabular}
\end{table*}

To determine the most appropriate modelling approach, we conducted an in-depth analysis of the dataset, focusing particularly on the statistics of CWEs and CAPECs. The dataset comprises a total of 1,067 records, of which 94.4\% contain only a single CWE, while 5.6\% include two CWEs (a negligible number of records contain up to five CWEs, which were excluded from further analysis due to their insignificance). Regarding the distribution across Advanced Persistent Threat (APT)-CAPEC categories (CAPEC labels), we observed the following: State Attack and Brute Force each constitute 7.5\% of the dataset, Social Engineering and Information Gathering both account for 28.7\%, Function Abuse represents 48.2\%, Illegal Access 58.4\%, Injection 62.4\%, and Data Manipulation 69.7\%. Based on these statistics, we concluded that a multi-class, multi-label classifier would be most suitable for predicting CAPECs. CWEs are associated with one to eight CAPECs, as detailed in Table \ref{table:CAPEC-DS-Distribution}. The dataset analysis revealed a significant imbalance in the data distribution. Rather than modifying the dataset, we addressed this imbalance by leveraging the built-in weighting mechanism of the Gradient Boosting Machine (GBM) model, which adjusts classification probabilities to handle under-represented labels better.

We applied the same data transformation techniques as outlined in \cite{DBLP:conf/mobiquitous/AlvarezSNK21}, including label encoding for categorical features, text frequency normalisation for numerical values, and data scaling. Additionally, we reduced data sparseness using the standardisation function.

\begin{table}
\caption{Dataset Labels Distribution}
\label{table:CAPEC-DS-Distribution}
\centering
\begin{tabular}
{|m{2.5cm} |m{1.5cm} |m{1.8cm}|}
\hline
\textbf{Associated Labels} &\textbf{Counts} &\textbf{Distribution \%}\\
\hline 
1 CAPECs &320      &30\%\\
\hline 
2 CAPECs &1        &0.1\%\\
\hline 
3 CAPECs &434      &40.7\%\\
\hline 
4 CAPECs &0        &0\%\\
\hline 
5 CAPECs &226      &21.2\%\\
\hline 
6 CAPECs &0        &0\%\\
\hline 
7 CAPECs &75       &7\%\\
\hline 
8 CAPECs &5        &0.5\%\\
\hline 
\end{tabular}
\end{table}

\subsection{Model Selection}

To select the type of machine learning classification architecture for CAPEC prediction in IoT devices, we consider the following factors:
\begin{itemize}
\item{Factor-1: The dataset (content, structure, and size).}
\item{Factor-2: The Machine Learning algorithms trained with this dataset.}
\item{Factor-3: The type of prediction intended.}
\end{itemize}

Our analysis for Factor-1 indicates that the dataset we will use is a construct that uses as a baseline the dataset we created in \cite{DBLP:conf/mobiquitous/AlvarezSNK21}. We integrate the CWEs and CAPECs into each record. For Factor-2, we take that the type of Machine Learning algorithm we used (Gradient Boosting Machine or GBM from sklearn\footnote{https://scikit-learn.org}) was effective. For comparison purposes, we implemented the Histogram-based Gradient Boosting Classification Tree, a Machine Learning algorithm derived from the GBM. Lastly, for Factor-3, we have determined the CAPECs to be the label to predict. Note that the CAPECs are different attack patterns that may occur simultaneously as single or multiple attacks; hence, the approach for this problem is a multi-class/multi-label classifier.

To evaluate the model's performance, we used two approaches oriented for multi-label classification \cite{DBLP:reference/dmkdh/TsoumakasKV10}. Firstly, the label-based evaluation utilises Precision, Recall, F1-score, and Accuracy metrics.

\begin{equation} \label{metrics}
\scalebox{0.85}{$
\begin{split}
Precision=\frac{1}{m}\sum_{i=1}^{m}\frac{\big| Y_{i} \cap Z_{i} \big|}{\big| Z_{i} \big|}
\quad Recall=\frac{1}{m}\sum_{i=1}^{m}\frac{\big| Y_{i} \cap Z_{i} \big|}{\big| Y_{i} \big|}\\
F1=\frac{1}{m}\sum_{i=1}^{m}\frac{2\big| Y_{i}\cap Z_{i} \big|}{\big|Z_{i}\big|+\big|Y_{i}\big|}
\quad Accuracy=\frac{1}{m}\sum_{i=1}^{m}\frac{\big| Y_{i} \cap Z_{i} \big|}{\big|Y_{i}\cup Z_{i}\big|}\\
\end{split}
$}
\end{equation}

Evaluating a multi-label classifier is done through macro-averaging (using the individual class labels and then averaging over all classes) and micro-averaging (using all instances and class labels). These calculations are used over Precision, Recall, F1-score and Accuracy metrics. We can use the operations of a typical binary evaluation; B($tp$,$tn$,$fp$,$fn$) is calculated based on the number of true positives ($tp$), true negatives ($tn$), false positives($fp$) and false negatives ($fn$). After binary evaluation for a label $\lambda$ $tp\lambda$, $pf\lambda$, $tn\lambda$, and $fn\lambda$ are the number of true positives, false positives, true negatives and false negatives, respectively. The macro-averaged and micro-averaged versions of B are calculated using the following equations:

\begin{equation} \label{macromicro}
\scalebox{0.85}{$
\begin{split}
B_{macro}=\frac{1}{q}\sum_{\lambda=1}^{q} B(tp\lambda,tn\lambda,fp\lambda,fn\lambda) \\
B_{micro}=B \big( \sum_{\lambda=1}^{q} tp\lambda, \sum_{\lambda=1}^{q}tn\lambda,\sum_{\lambda=1}^{q}fp\lambda,\sum_{\lambda=1}^{q}fn\lambda \big)
\end{split}
$}
\end{equation}

One characteristic of micro-averaging and macro-averaging is that the resulting value is the same for accuracy, while it will differ for precision and recall metrics.

From \cite{DBLP:reference/dmkdh/TsoumakasKV10, DBLP:journals/pr/BoutellLSB04}, we take our second multi-label evaluation metric approach, the Alpha-score. Following the formula score and parameters.

\begin{equation} \label{alpha}
\scalebox{0.85}{$
\begin{split}
score(P_{x}) = \left( 1- \frac{\big| \beta M_{x} + \gamma F_{x} \big|} {\big| Y_{x} \cup P_{x} \big|} \right)^\alpha \\ \\
(\alpha \geq 0 , 0 \leq \beta, \gamma \leq 1, \beta = 1 \big| \gamma = 1) 
\end{split}
$}
\end{equation}

Here, Y$_{x}$ is the set of ground truth labels for test data x, P$_{x}$ is the set of predicted labels by the CAPEC classifier h. Let M$_{x}$ represent the missed labels M$_{x}$=Y$_{x}$-P$_{x}$, and F$_{x}$ the false positive labels F$_{x}$=P$_{x}$-Y$_{x}$. The constraints $\beta$ and $\gamma$ limit the score to non-negative. These parameters will penalise the false positives and missed predictions differently. Further, their constraining behaviour allows the evaluation measure to be customised to the application as required. On the other hand, the $\alpha$ parameter (forgiveness rate or alpha score) reflects how much to forgive label prediction errors. Small values are considered aggressive (forgiving errors), while values bigger than one are considered conservative (error penalisation is harsher). When tending to the limits such as $\alpha$= $\infty$, the score P$_x$=1, only when the prediction is fully correct, and 0 otherwise (more conservative); and when $\alpha$=0, the score P$_x$=1 except when the answer is fully incorrect (most aggressive).

In multi-label classification, the alpha score is helpful as it evaluates the models' capability to predict all the labels correctly. This alpha-based score utilises three parameters, which are denoted by the Greek letters alpha (also known as the forgiveness rate), beta (constraint for missed labels) and gamma (constraint for false positive labels). In \cite{DBLP:reference/dmkdh/TsoumakasKV10, DBLP:journals/pr/BoutellLSB04}, the authors tested different values for the beta and gamma parameters, and in their results, it was found that setting beta and gamma with the same value would force the metric to penalise both false positives and misses equally and give all the weight of the score to the alpha parameter. Setting alpha with small values (close to zero), the metric will tend to forgive errors, and setting alpha with significant values (equal to or bigger than 1), the metric will penalise errors more harshly. Hence, the metric will return one only when the prediction is fully correct, zero when the prediction is fully incorrect (missed), and a penalised result otherwise (partly-correct). For instance, our model receives a dataset that predicts up to eight CAPECs. With the alpha score, we can measure the correctness of the label's prediction. 
When measuring labels prediction, we will have fully-correct, partly-correct, or fully-incorrect (missed). 
Using the example of the device with eight labels, we would label the prediction as follows:
\begin{itemize}
\item{The prediction is fully-correct if the eight labels are correctly predicted 
{(For example: 8 correct - 0 incorrect).}}
\item{The prediction is partly-correct in the cases where at least one of the predicted labels is correct 
{(For example: 1 correct - 7 incorrect, 2 correct - 6 incorrect ... 7 correct - 1 incorrect).}} 
\item{The prediction is fully-incorrect if none of the predicted labels corresponds to the true labels 
{(For example: 0 correct - 8 incorrect).}}
\end{itemize}

The performance results of the three multi-label classifiers GBM, XGBM, and RF are disclosed in Table \ref{table:CAPECs_performance}.
We conclude that:
\begin{itemize}
\item{The three models performed similarly; although we used the XGBM and RF to benchmark, the XGBM achieved slightly better results.}
\item{The alpha scores confirm that any model (GBM, XGBM or RF) can predict the CAPECs associated with IoT devices.} 
\item{The alpha score as 0 (forgiving errors) returned for all three models a value of one, meaning the prediction is fully correct.} 
\item{The alpha score as 1 (penalise errors harshly) returned 97.4\% as a result of the RFC model and 99.4\% for GBMC and XGBX models. This means the models are performing partly correct (1 prediction correct and 7 predictions incorrect).} 
\end{itemize}

\begin{table}
\caption{Performance results from the multi-label classification models GBM, XGBM, and RF.}
\label{table:CAPECs_performance}
\centering
\begin{tabular}
{|m{2cm}      |m{1.5cm}          | m{1.5cm} | m{1.6cm} |}
\hline
\textbf{Metric} &\textbf{GBMC} &\textbf{XGBC}& \textbf{RFC}\\
\hline 
AV Accuracy         & 99.4\%     &99.4\%     &96.7\%  \\
\hline
Macro Accuracy      & 98.9\%     &99.3\%     &94.4\%  \\
\hline
Micro Accuracy      & 99.4\%     &99.5\%     &97.4\%  \\
\hline
Macro Precision     & 99.5\%     &99.9\%     &99.3\%  \\
\hline
Micro Precision     & 99.7\%     &99.8\%     &99.5\%  \\
\hline
Macro Recall        & 99.4\%     &99.5\%     &95.0\%  \\
\hline
Micro Recall        & 99.6\%     &99.7\%     &98.0\%  \\
\hline
Micro F1            & 99.6\%     &99.7\%     &96.9\%  \\
\hline
Macro F1            & 99.6\%     &99.7\%     &98.7\%  \\
\hline
Alpha Score(0)      &100.0\%     &100.0\%    &100.0\%      \\
\hline
Alpha Score(1)      & 99.4\%     &99.5\%     &97.4\%  \\
\hline
\end{tabular}
\end{table}

\subsection{Experimental Setup and Evaluation for Attack Patterns Prediction Framework}
Our experimental setup was implemented over the paid Google Colab Pro+ platform. The execution engine works under Python version 3.0 with Sklearn, Pandas, and Numpy libraries. The resources are allocated as a default configuration with variable amounts such as 0.8-50 GB in RAM and 38-107 GB in Hard Drive space.

We used the best values for the parameters from \cite{DBLP:conf/mobiquitous/AlvarezSNK21}. We list these values as follows: 10,000 estimators (number of boosting stages to perform), a learning rate of 0.01 (shrinks the contribution of each tree), 500 as the maximum depth (of the individual regression estimators), and 1e-2 as the minimum of impurity decrease (A node will be split based on this parameter). These parameters were employed as equally as possible (some are exclusive of the GBM) in the three classification models: GBM, XGBM and RF. Table \ref{table:CAPECs_performance} presents the result of our performance analysis confirming an accuracy of 99.4\% with GBM model.

\section{Conclusion and Future Work}
We present a comprehensive platform designed to predict weaknesses and attack patterns in IoT devices. This platform is particularly valuable for pre-deployment security checks, which are traditionally costly, time-consuming, and often require physical access to the device. To support this platform, we have curated a uniquely balanced dataset specifically for IoT security, integrating information from multiple authoritative sources, including Common Vulnerabilities and Exposures (CVE), Common Weakness Enumeration (CWE), Common Attack Pattern Enumeration and Classification (CAPEC), as well as other publicly accessible online data repositories. In developing our platform, we conducted an extensive review of state-of-the-art machine learning approaches. Based on this review, we fine-tuned and implemented a Bidirectional Long Short-Term Memory (Bi-LSTM) model, achieving 77\% accuracy in predicting potential weaknesses associated with vulnerabilities in IoT devices. Additionally, by applying a Gradient Boosting Machine (GBM) model, we achieved an impressive 99.4\% accuracy in detecting attack patterns linked to the identified weaknesses. While our current solution is optimised for IoT devices with network connectivity, future work will aim to diversify the types of IoT devices included in our dataset. This will involve expanding our focus to include devices connected through bridge networks or those that operate using restricted protocols. 

\section{Dataset}

The datasets used in this paper are publicly available at: \url{https://github.com/criveraalvarez/IoT_CWE-CAPEC_Dataset}.

\bibliographystyle{ieeetr}
\bibliography{ref}

\end{document}